\documentclass[aps,twocolumn,showpacs]{revtex4}
\usepackage{color}
\usepackage{graphicx}
\usepackage{amsmath}
\usepackage{amssymb}
\usepackage{amsfonts}
\usepackage[colorlinks,hyperindex]{hyperref}
\usepackage{epsfig}

\def\beq{\begin{equation}}
\def\eeq{\end{equation}}
\def\nR{n_{\rm R}}

\begin{document}

\title{Adiabatic approximation and fluctuations in exciton-polariton condensates}
\author{Nataliya Bobrovska and Micha{\l} Matuszewski}
\affiliation{
Institute of Physics Polish Academy of Sciences, Al. Lotnik\'ow 32/46, 02-668 Warsaw, Poland}

\begin{abstract}
We study the relation between the models commonly used to describe the dynamics of nonresonantly pumped 
exciton-polariton condensates, namely the 
ones described by the complex Ginzburg-Landau equation, and by the open-dissipative Gross-Pitaevskii equation 
including a separate equation for the reservoir density.
In particular, we focus on the validity of the adiabatic approximation that allows to reduce the coupled condensate-reservoir
dynamics to a single partial differential equation. 
We find that the adiabatic approximation consists of three independent analytical conditions that have to be fulfilled 
simultaneously. By investigating stochastic versions of the two corresponding models, 
we verify that the breakdown of the adiabatic approximation can lead to discrepancies in correlation lengths and 
distributions of fluctuations. Additionally, we consider the phase diffusion and number fluctuations
of a condensate in a box, and show that self-consistent description requires treatment beyond the typical Bogoliubov approximation.
\end{abstract}

\pacs{67.85.De, 71.36.+c, 03.75.Kk}

\maketitle

\section{Introduction}

Exciton-polaritons are coherent superpositions of quantum well excitons and a microcavity photons,
resulting from strong coupling of the two modes at resonance~\cite{Hopfield_Polaritons,Weisbuch_Polaritons,Kavokin_Microcavities}. 
The mixed nature of these quasiparticles is attractive from the point of view of fundamental and applied 
research in many ways. The matter component provides strong interactions, while the photonic
component yields very light effective mass and allows for straightforward detection.
Condensation and supefluidity of polaritons or polariton lasing has been demonstrated
in many laboratories, even at room temperature~\cite{Amo_Superfluidity,Deveaud_Vortices,Carusotto_QuantumFluids,
Kasprzak_BEC,Grandjean_RoomTempLasing,Kena_Organic,Mahrt_RTCondensatePolymer,Yamamoto_RMP}.

In the case of nonresonant pumping, polaritons can be created either using a beam at frequency above the
polariton resonance or using electrical carrier injection~\cite{Hofling_ElectricallyPumped}. Free carriers and high-energy
excitons undergo energy relaxation towards the polariton ground state, where they can condense~\cite{Kasprzak_BEC}.
Theoretical description of this complicated process, involving scattering with phonons and other polaritons 
is a formidable task, and several approaches have been proposed in the past
with various approximations involved~\cite{Haug_QuantumKineticGPDerivation,Wouters_ClassicalFields,Malpuech_Hybrid,Laussy_SpontaneousCoherence,Bloch_Molecules,Tassone_Bottleneck}. Among these,
phenomenological models based on various generalizations of the Gross-Pitaevskii 
equation~\cite{Keeling_VortexDynamics,Wouters_Excitations,Wouters_Energyrelaxation}
have been particularly useful thanks to their simplicity and limited number of external parameters. 
The most commonly used are the complex Ginzburg-Landau equation (CGLE, also termed the generalized GP equation in some works), 
with a single equation for the condensate evolution,
and the open-dissipative Gross-Pitaevskii equation (ODGPE, other names are also used in the literature)~\cite{Wouters_Excitations}, 
which attempts to describe the full system dynamics using a pair of coupled condensate-reservoir equations.
It has been pointed out in several places~\cite{Dreismann_Counterrotating,Berloff_Universality,Altman_DrivenSuperfluid2D} that the 
ODGPE model can be reduced to the CGLE model under the adiabatic (quickly responding reservoir) approximation, but the 
validity of this approximation has not been investigated in detail.

In this paper, we investigate systematically the relation between the the CGLE
the ODGPE models.
We establish precisely conditions under which the reduction to the CGLE model is justified;
contrary to the common belief, we show that the fast reservoir relaxation alone is not a sufficient condition.
We show that adiabaticity requires three independent analytical conditions to be fulfilled simultaneously.
Additionally, the condensate must remain close to the steady state, since large fluctuations may lead to 
complete breakdown of the correspondence between the models. Such large fluctuations occur in particular
close to the condensation/stability limits of the condensate phase diagram.

In the second part of the paper, we investigate how the breakdown of the adiabaticity influences 
the steady state solutions of the corresponding stochastic CGLE/ODGPE models. 
Recently, fluctuations of nonequilibrium quantum fluids 
became a very active area of research. Spatial~\cite{Carusotto_NonequilibriumQuasicondensates,Wouters_SpatialCoherence} 
and temporal~\cite{Wouters_TemporalCoherence,Diehl_ScalingProperties} correlations have been investigated 
in the small fluctuations regime, and the critical scaling properties have been established using the
renormalization group~\cite{Sieberer_DynamicalCritical,Sieberer_NonequilibriumRG,Diehl_ScalingProperties}. Dynamics of the polariton condensation phase transition
have been shown to display similarities with the Kibble-Zurek theory of universal dynamics~\cite{Matuszewski_UniversalityPolaritons,Liew_InstabilityInduced}.
Here, we show how the spectrum of fluctuations and spatial correlations are modified in the non-adiabatic
regime as well as in the large fluctuations regime. In particular, we demonstrate the appearance of dark soliton-like
structures and chaotic bistable steady states close to the limits of condensate stability in parameter space in the ODGPE model.

In addition, we consider a model of a condensate in a box, 
in which case condensation may occur despite the absence of true long-range order 
at low dimensions~\cite{Mora_Quasicondensates, Shlyapnikov2004, Carusotto_NonequilibriumQuasicondensates}. 
We show that a self-consistent description requires treatment beyond the typical Bogoliubov approximation, and demonstrate 
how the zero-momentum singularity of the momentum distribution spectrum~\cite{Carusotto_NonequilibriumQuasicondensates} can be avoided by
an appropriate generalization of the Bogoliubov ansatz. This allows for the determination of
the number fluctuations and condensate phase diffusion equation.

The paper is organized as follows. In Sec.~\ref{sec:models} we introduce the CGLE and ODGPE models as well as their stochastic versions,
in particular the stochastic Gross-Pitaevskii (SGPE) model being a generalization of the CGLE. 
In Sec.~\ref{sec:adiabatic} we derive carefully the conditions under which the adiabatic reduction of the ODGPE to the CGLE
model is justified. In Sec.~\ref{sec:fluctuations} we recall briefly the main analytical results concerning the
fluctuations and spatial correlations of the condensate. We also present the analysis of the fluctuations
and phase diffusion of a condensate confined in a box. In Sec.~\ref{sec:numerical} we present numerical investigation
of the properties of stochastic steady states in various regimes. Sec.~\ref{sec:conclusions} concludes the paper.

\section{Models} \label{sec:models}

\subsection{Complex Ginzburg-Landau equation}

The simplest dissipative model to describe one-dimensional exciton-polariton condensates is the
complex-Ginzburg-Landau equation (CGLE)
\begin{align} \label{CGLE}
i\textrm{d}\phi=&\left[A-B\nabla^2+C|\phi|^2+i\left(D-E|\phi|^2\right)\right]\phi \,{\rm d}t + \nonumber \\
+& {\rm d}W_{\rm CGLE},
\end{align}
where for future reference we additionally included the stochastic term ${\rm d}W_{\rm CGLE}({\bf x},t)$ which vanishes in the classical limit. 
The real parameters of the equation describe the energy offset ($A$), the dispersion coefficient ($B$), nonlinear interactions ($C$), 
external pumping ($D$), and nonlinear losses ($E$). In the case of interest the parameters $B$, $C$, $D$ and $E$ are positive,
which guarantees the existence of a stable homogeneous solution $\phi({\bf x},t)=\phi_0e^{-i\mu t}=\sqrt{D/E}e^{-i\mu t}$ 
with $\mu=A+C|\phi_0|^2$ ~\cite{Hohenberg_FrontsPulses}.
Compared to the most general form of the CGLE, we assume no diffusive term (as $B$ is real), although models including 
the diffusive term have also been employed~\cite{Bloch_1DAmplification,Keeling_VortexDynamics,Sieberer_DynamicalCritical}. It is, however, 
not crucial to our considerations. We note that the coefficient
$A$ can be removed simply by moving to a rotating frame where $\phi'({\bf x},t) = \phi({\bf x},t)e^{iAt}$, provided that the noise 
is not time-correlated.

A generalized form of the CGLE, or the stochastic Gross-Pitaevskii equation (SGPE), has been used in several works which investigated 
spatial correlations of the condensate \cite{Wouters_Excitations, Carusotto_NonequilibriumQuasicondensates}
\begin{eqnarray}
\label{SGPE}
i\textrm{d}\phi&=&\left[\omega_0-\frac{\hbar}{2m_0^*}\nabla^2+g_0|\phi|^2\right.+\\\nonumber
&+&\left.i\left(\frac{P_0}{1+\frac{|\phi|^2}{n_s}}-\gamma_0\right)\right]\phi \,\textrm{d}t+\textrm{d}W_{\rm SGPE},
\end{eqnarray}
where $\phi({\bf x},t)$ is the wave function, 
$m_0^*$ is the effective mass of lower polaritons,
$g_0$ is the interaction coefficient, 
$\omega_0$ is the oscillation frequency,
$P_0$ is the pumping rate,
$\gamma_0$ is the polariton loss rate, $n_s$ is the saturation density, and 
$\textrm{d}W_{\rm SGPE}({\bf x},t)$ is the complex stochastic noise.

As in the second part of the manuscript we will focus on the spatial correlations, we will compare our results
to the above model~(\ref{SGPE}), which was used in~\cite{Wouters_Excitations, Carusotto_NonequilibriumQuasicondensates}, rather than the standard CGLE~(\ref{CGLE}). 
These two models are, however, completely equivalent
as long as one is interested in a state with small density fluctuations around the homogeneous solution. To show this
consider the solution for which the density can be written as $|\phi({\bf x},t)|^2=|\phi_0|^2 + \delta n$, where $\delta n / |\phi_0|^2\ll 1$.
The fraction in Eq.~(\ref{SGPE}) can be expanded in Taylor series, and by neglecting second- and higher-order terms in $\delta n$
we obtain Eq.~(\ref{CGLE}) with the correspondence $A=\omega_0$, $B=\hbar/2m_0^*$, $C=g_0$, $D=\gamma_0-(\gamma_0^2/P_0)$,
$E=\gamma_0^2/n_sP_0$, $\textrm{d}W_{\rm CGLE}=\textrm{d}W_{\rm SGPE}$, which yields the steady state density $|\phi_0|^2=n_s(P_0-\gamma_0)/\gamma_0$ for $P_0>\gamma_0$.

For completeness, we note that in~\cite{Wouters_Excitations, Carusotto_NonequilibriumQuasicondensates} the stochastic noise was assumed to be Gaussian with the correlations
\begin{align}
\label{noise-phi}
\langle\textrm{d}W_{\rm SGPE}({\bf x})\textrm{d}W_{\rm SGPE}^*({\bf x}^{\prime})\rangle&=2D_{\phi\phi}\delta({\bf x}-{\bf x}^{\prime})\textrm{d}t,\nonumber\\
\langle\textrm{d}W_{\rm SGPE}({\bf x})\textrm{d}W_{\rm SGPE}({\bf x}^{\prime})\rangle&=0
\end{align}
and the amplitude $D_{\phi\phi}\approx\gamma_0$, which is an estimate of the quantum noise due to the dissipative coupling.
Since the above white noise $\textrm{d}W_{\rm SGPE}$ has a diverging norm, in practice an appropriate UV cutoff must be employed
eg.~through frequency dependence of the amplification term~\cite{Carusotto_NonequilibriumQuasicondensates}.
Alternatively, one may model the process on a discretized mesh with lattice constant $\Delta x$, 
where the Dirac delta is replaced as $\delta({\bf x}-{\bf x}')\rightarrow \delta_{{\bf x},{\bf x}'}/(\Delta x)^d$.
We note that an alternative form of stochatic noise was derived from a quantum model of condensation
in a cavity interacting with two-level emitters~\cite{carusotto_Langevin}.


\subsection{Open-dissipative Gross-Pitaevskii equation}

A more realistic model of the polariton condensate, called the open-dissipative Gross-Pitaevskii equation (ODGPE), 
includes coupling of the condensate wave function to the reservoir described
by a density field $\nR({\bf x},t)$
~\cite{Wouters_Excitations, Wouters_ClassicalFields}
\begin{align}
\label{GPE-psi}
i\textrm{d}\psi=&\left[ -\frac{\hbar}{2m^*}\nabla^2+g_C|\psi|^2+g_Rn_R+\right.\\\nonumber
&\left.+\frac{i}{2}\left(Rn_R-\gamma_C\right)
\right]\psi\textrm{d}t+\textrm{d}W,\\
\label{GPE-nr}
\frac{\partial n_R}{\partial t}=&P-\left(\gamma_R+R|\psi|^2\right)n_R,
\end{align}
where the complex stochastic noise can be obtained within the truncated Wigner approximation~\cite{Wouters_ClassicalFields}
\begin{align}
\label{noise-psi}
\langle\textrm{d}W({\bf x})\textrm{d}W^*({\bf x}^{\prime})\rangle&=\frac{\textrm{d}t}{2(\Delta  x)^d}(Rn_R+\gamma_C)\delta_{{\bf x},{\bf x}^{\prime}}
,\\
\langle\textrm{d}W({\bf x})\textrm{d}W({\bf x}^{\prime})\rangle&=0.
\end{align}
where we assumed a constant scattering rate $R$ which gives noise correlations analogous as in Eq.~(\ref{noise-phi}).
Here $P({\bf x})$ is the exciton creation rate determined by the pumping profile,
$\gamma_{\rm C}$ and $\gamma_{\rm R}$ are the polariton and reservoir loss rates, 
and $R$ is the rate of stimulated scattering from the reservoir to the condensate, and $g_{\rm C}$, $g_{\rm R}$ are
the rates of repulsive polariton-polariton and reservoir-polariton interactions, respectively.

In the absence of noise, a spatially uniform solution is given by $\psi({\bf x},t)=\psi_0e^{-i\mu_0t}$, $n_R({\bf x},t)=n_R^0$.
Above the threshold pumping $P>P_{\rm th}=\gamma_C\gamma_R/R$ a stable condensate exists with the condensate density $|\psi_0|^2=(P/\gamma_C)-(\gamma_R/R)$, $n_R^0=\gamma_C/R$, and $\mu_0=g_C|\psi_0|^2+g_Rn_R^0$. We note that formally similar models were also used to describe
nonlinear effects in semiconductor microcavities at weak coupling~\cite{Taranenko2005}
and atom lasers~\cite{Kneer_GenericAtomLaser}.
The above phenomenological model has been successful in describing a number of 
different experimental situations in exciton-polariton condensates, although 
various values of parameters have been used in the 
literature~\cite{Bobrovska_Stability,Bloch_1DAmplification,Deveaud_VortexDynamics,Yamamoto_VortexPair}.

\section{Adiabatic approximation and the correspondence between the models}  \label{sec:adiabatic}

\subsection{Adiabatic limit}

The open-dissipative Gross-Pitaevskii model Eqs.~(\ref{GPE-psi})-(\ref{GPE-nr}) can be reduced to the simpler and more tractable CGLE 
model Eq.~(\ref{CGLE}) under the assumption that the reservoir density $\nR({\bf x},t)$ adiabatically follows the change of $|\psi({\bf x},t)|^2$.
With the use of Eq.~(\ref{GPE-nr}) we can express $\nR$ as
\beq
\nR({\bf x},t)=\frac{P}{\gamma_R + R |\psi({\bf x},t)|^2}.
\eeq
Note that the above relation is valid also in the case of a stationary state $\nR({\bf x},t)=\nR({\bf x})$, $\psi({\bf x},t)=\psi({\bf x})e^{-i\mu t}$, 
regardless of the adiabatic assumption. However, in this work we are interested in dynamical processes and the limits
of validity of the adiabatic approximation.
The Eq.~(\ref{GPE-psi}) now takes the form of a generalized CGLE
\begin{eqnarray}
\label{CGLE-nr}
i\textrm{d}\psi&=&\left[-\frac{\hbar}{2m_0}\nabla^2+g_C|\psi|^2
+\frac{g_R P}{\gamma_R + R |\psi({\bf x},t)|^2}\right.+\nonumber\\
&+&\left.\frac{i}{2}\left(\frac{RP}{\gamma_R + R |\psi({\bf x},t)|^2}-\gamma_C\right)\right]\psi \,\textrm{d}t+\textrm{d}W.
\end{eqnarray}

Again, this equation is equivalent to Eqs.~(\ref{CGLE}) and~(\ref{SGPE}), provided that the condensate density $|\psi({\bf x},t)|^2$ 
is close to the homogeneous solution $|\psi_0|^2=(P/\gamma_C)-(\gamma_R/R)$ which allows for expansion of the right hand side 
in Taylor series
to the first order. We emphasize that the expansion is done around the steady state density, and not zero density of the condensate. As
will be shown in subsection~\ref{sec:large_fluctuations}, at low density (pumping slightly above threshold) 
the results of the two models disagree qualitatively 
due to large fluctuations.
The full correspondence between the three models is given by 
\begin{align}
&A=\omega_0=g_R\left(\frac{2\gamma_C}{R}-\frac{\gamma_C^2\gamma_R}{R^2P}\right),\nonumber\\
&B=\frac{\hbar}{2m_0^*}=\frac{\hbar}{2m^*},\nonumber\\
&C=g_0=g_C-g_R\frac{\gamma_C^2}{RP},\label{correspondence} \\
&D=\gamma_0-\frac{\gamma_0^2}{P_0}=\frac{\gamma_C}{2}-\frac{\gamma_C^2\gamma_R}{2PR},\nonumber\\
&E=\frac{\gamma_0^2}{n_sP_0}=\frac{\gamma_C^2}{2P}, \nonumber\\
&D_{\phi\phi} =\gamma_0 = \gamma_C / 2.\nonumber
\end{align}
Note that there are only six equations while seven parameters are present in the 
open-dissipative model. Consequently, one free parameter can be chosen when calculating seven parameters of 
Eqs.~(\ref{GPE-psi})-(\ref{GPE-nr}) corresponding to six parameters of Eq.~(\ref{CGLE}) or Eq.~(\ref{SGPE}).
In the classical limit (no stochastic noise) the last equation is absent and there is an additional free parameter
in both Eqs.~(\ref{GPE-psi})-(\ref{GPE-nr}) and Eq.~(\ref{SGPE}).

\subsection{Validity of the adiabatic approximation}

The limits of the validity of the adiabatic approximation can be estimated by comparing the characteristic timescales
existing in the dynamical system. To this end consider Eqs.~(\ref{GPE-psi})-(\ref{GPE-nr}) and assume 
that the adiabatic approximation is fulfilled; the reservoir density $\nR$ is able to quickly adjust 
to the condensate density $|\psi|^2$ which changes on a much longer timescale.
Equation (\ref{GPE-nr}) has a simple solution if $|\psi({\bf x},t)|^2$ is treated as a constant
\beq
\nR({\bf x},t)=\frac{P}{\gamma_R + R |\psi({\bf x},t)|^2} + C e^{-(\gamma_R + R |\psi({\bf x},t)|^2)t}.
\eeq
Therefore the timescale of the reservoir is $\tau_R=(\gamma_R + R |\psi({\bf x},t)|^2)^{-1}$. For the adiabatic assumption to be consistent,
this timescale must be much smaller than the timescales of the condensate, which are given by all the four terms in
Eq.~(\ref{GPE-psi}). We obtain four conditions that, if fulfilled {\it simultaneously},
give a sufficient condition for the validity of the adiabatic approximation; namely, all the terms
$\frac{\hbar k^2}{2m^*}$, $\frac{1}{2}|Rn_R-\gamma_C|$, $g_C|\psi|^2$, and $g_Rn_R$ must be much smaller than $\tau_R^{-1}=\gamma_R+R|\psi|^2$.
The first condition means that only the low momentum modes of the condensate
\beq \label{condk}
k^2 \ll 2m^*/(\hbar\tau_R)
\eeq
may be occupied considerably. 
The other conditions provide relations between the system parameters, as we show below.

We now make the same assumption that was used to compare the different models of the condensate,
expressing the condensate density as a steady state value plus small fluctuations, 
$|\phi({\bf x},t)|^2=|\phi_0|^2 + \delta n$. Under this assumption, the second of the above conditions is always fulfilled. 
The third and fourth conditions take the simple forms
\begin{align}
\frac{P_{\rm th}}{P}\gg\frac{g_C-R}{g_C},\label{cond1}\\
\frac{P}{P_{\rm th}}\gg\frac{g_R}{R}\frac{\gamma_C}{\gamma_R}.\label{cond2}
\end{align} 
Equation~(\ref{cond1}) is always true if $g<R$, and otherwise it gives an upper limit for the pumping $P$. Equation~(\ref{cond2})
gives a lower limit for $P$. 
Note that the condition~(\ref{cond1}) is {\it independent} of the reservoir relaxation rate $\gamma_R$.

Finally, we note that while the above conditions are sufficient, but not necessary, it is unlikely that if one of the four terms is large,
it could be fully compensated by another term in Eq.~(\ref{GPE-psi}), since the terms depend very differently on the fields 
$\nR$ and $\psi$. We also emphasize once again that Eqs.~(\ref{cond1}) and~(\ref{cond2})  as well as the exact correspondence
between the models are only valid if the condensate density is close to the steady state value.

\subsection{Limits of modulational stability}

It is well known that the CGLE~(\ref{CGLE}) displays 
linear modulational instability (also called Benjamin-Feir instability) for 
$BC<0$, which is a special case of the Benjamin-Feir-Newell criterion~\cite{Aranson_CGLEWorld,Hohenberg_FrontsPulses}. 
While the dispersion coefficient $B$ is always positive in our model, it follows from Eqs.~(\ref{correspondence}) 
that not all choices of the interaction coefficients $g_C>0$ and $g_R>0$ correspond to a positive value of $C$.
A simple application of Eqs.~(\ref{correspondence}) yields the criterion for linear stability
\begin{equation}
\label{condition}
\frac{P}{P_{\rm th}}>\frac{g_R}{g_C}\frac{\gamma_C}{\gamma_R}.
\end{equation}
In other words, the violation of the stability condition~(\ref{condition}) corresponds to effectively attractive
interactions in the CGLE or SGPE model, $C,g_0<0$.
The above instability of the model~(\ref{GPE-psi})-(\ref{GPE-nr}) was first reported in~\cite{Wouters_Excitations}, 
where it was named the ``hole-burning effect'', and the analytical condition for stability~(\ref{condition}) 
was first derived in~\cite{Ostrovskaya_DarkSolitons} by investigation of the Bogoliubov-de Gennes excitation spectrum. 
Alternatively, the same condition can be derived from the analysis of the total blueshift originating from both $g_C$ and 
$g_R$~\cite{Liew_InstabilityInduced}.


\section{Fluctuations and spatial correlations} \label{sec:fluctuations}

In the rest of the paper, we investigate how the breakdown of the adiabatic approximation influences the fluctuations
around the steady state and spatial correlation functions. 
We first recall the relevant results of~\cite{Carusotto_NonequilibriumQuasicondensates} and~\cite{Wouters_SpatilaCorelations}
corresponding to the SGPE (or adiabatic) case.
We also derive self-consistently analytical formulas for the number fluctuations and phase diffusion of a condensate in a finite box.
In the last section we present numerical results in both the adiabatic and non-adiabatic regimes.

\subsection{Previous analytical results}

In~\cite{Carusotto_NonequilibriumQuasicondensates} Bogoliubov-de Gennes equations for the small fluctuations around the condensate
in the form $\phi({\bf x},t)=\left[\phi_0+ \delta \phi({\bf x},t)\right]e^{-i\mu t}$ were used to determine the momentum
distribution  of the particles in the SGPE model~(\ref{SGPE}). The number of particles in each mode in the steady state was calculated as
\begin{equation}
\label{tail}
n^{ss}_{\textbf{k}}=(2\pi)^d |\phi_0|^2 \delta^{(d)}(\textbf{k}) + \frac{D_{\phi\phi}}{\Gamma}\left[\frac{\mu^2+\Gamma^2}{E^2_{\textbf{k}}}+1\right],
\end{equation}
where $\Gamma=\gamma_0(P_0-\gamma_0)/P_0$ is the effective damping rate, $\mu=g_0|\phi_0|^2>0$ is the energy of 
interactions between particles, and $E^2_{\textbf{k}}=\epsilon_{\textbf{k}}(\epsilon_{\textbf{k}}+2\mu)$ 
with $\epsilon_{\textbf{k}}=\hbar k^2/2m^*$ is the Bogoliubov dispersion.
The first term on the right hand side corresponds to the condensate, and the second term to the non-condensed cloud.

The formula~(\ref{tail}) must be used with caution. It displays both the IR divergence in the $E^{-2}_{\textbf{k}}$ term
and the UV divergence in the constant term. While the UV divergence can be healed in the case of frequency-dependent 
pumping~\cite{Carusotto_NonequilibriumQuasicondensates} or discretization of the system on a lattice, 
the IR divergence poses a problem for the interpretation of the above formula 
at the $k=0$ point. We show below how this problem can be solved by an appropriate generalization of the Bogoliubov approximation.

Notice also that in the infinite system (``thermodynamic'' limit with $V\rightarrow \infty$ at constant pumping) the
fraction of particles in the condensate vanishes in the relevant one- and two-dimensional cases~\cite{Carusotto_NonequilibriumQuasicondensates},
invalidating the Bogoliubov approximation. However, the long-range correlations are mostly influenced by the
phase fluctuations, which are much less energetically costly than the amplitude fluctuations due to the existence of a Goldstone mode
of phase twists~\cite{Mora_Quasicondensates, Shlyapnikov2004}. 
The correlations are accurately described using the density-phase generalized Bogoliubov approximation
$\phi({\bf x},t)=\sqrt{n_0+\delta n} \,e^{i\theta({\bf x},t)-i\mu t}$, which yields the long-range correlations
in one dimension falling exponentially as~\cite{Carusotto_NonequilibriumQuasicondensates, Wouters_SpatilaCorelations}
\beq \label{g1}
g^{(1)}(x) \sim \exp(-x/l_1), \quad {\rm where} \quad l_1=\frac{2 n_0 \hbar \mu \Gamma}{m D_{\phi\phi} (\mu^2+\Gamma^2)}.
\eeq

\subsection{Condensate in a box}

In a finite system where ${\bf k}$ is discretized, the fraction of
particles in the condensate $k=0$ can be large, and we will consider this case in this subsection. 
In the following we consider a condensate in a box of length $L$ and periodic boundary conditions with $\Delta k=2\pi/L$, for which we 
have the correspondence $\delta^{(d)}(\textbf{k})\rightarrow (L/(2\pi))^d \delta_{k,0} $. The first term on the right hand side 
of~(\ref{tail}) can be written simply as $(2\pi)^d |\phi_0|^2 \delta^{(d)}(\textbf{k}) = |\phi_0|^2 L^d\delta_{k,0} \equiv N_0\delta_{k,0}$.
To cope with the IR divergence due to the singularity at $k=0$ we introduce a generalized density-phase Bogoliubov Ansatz 
which takes into account the diffusion of the condensate phase $\varphi(t)$
\beq
\phi({\bf x},t)=\left[\phi_0+\delta\phi_{k= 0}(t)\right] e^{i\varphi(t)-i\omega t} +\delta\phi_{k\neq 0}({\bf x},t),
\eeq 
where $\varphi(t)$ and $\delta\phi_{k= 0}(t)$ are real functions describing the phase and amplitude fluctuations of the condensate, 
and $\delta\phi_{k\neq 0}({\bf x},t)$ is a complex function describing the fluctuations in the out-of-condensate modes.
In contrast to the standard Bogoliubov approach, we treat the fluctuations of the $k=0$ mode in a special way. We do not assume that the
phase $\varphi(t)$ is a small quantity, but allow it to undergo a Brownian-like motion under the action of 
the stochastic noise~\cite{ScullyZubairy}.
Substituting the above into Eq.~(\ref{SGPE}) and going into Fourier space, the fluctuations $\delta\tilde{\phi}_{k\neq 0}(k,t)$ separate
from $\delta\tilde{\phi}_{k= 0}(t)$ and $\varphi(t)$ in the linearized limit.
One gets a set of equations
\begin{align}
\label{im_noise}
\frac{d\,\delta\phi_{k=0}}{dt}&=-2\Gamma\,\delta\phi_{k=0}+\frac{1}{L^{d/2}}\frac{dW_{I,k=0}}{dt},\\
\label{re_noise}
\frac{d\varphi}{dt}&=-\frac{1}{\sqrt{N_c}}\left(2\mu L^{d/2}\,\delta\phi_{k=0}+ \frac{dW_{R,k=0}}{dt}\right),
\end{align}
where $N_c(t)=L^d|\phi_0+\delta \phi_{k=0}|^2$ and we split the noise into the real and imaginary part 
with $\langle dW_{R,I,k=0} dW^*_{R,I,k=0} \rangle = D_{\phi\phi} dt$.
Equation~(\ref{im_noise}) corresponds to the Ornstein-Uhlenbeck process while Eq.~(\ref{re_noise}) 
describes phase diffusion. By stochastic integration of the first equation~\cite{Gardiner}  
we calculate the number fluctuations in the condensate 
\begin{equation}
\Delta n_{k = 0}=2N_0 \sqrt{\frac{\langle\delta\phi_{k=0}^2\rangle}{\phi_0^2}} = \sqrt{\frac{D_{\phi\phi} N_0}{\Gamma}}.
\end{equation}
As usual for an interacting condensate, the number fluctuations scale with the square root of $N_0$. 
The mean number of particles in the condensate and out-of-condensate modes  is given by (recall that $N_0 = |\phi_0|^2 L^d$)
\begin{equation}\label{nasz}
n^{ss}_{\textbf{k}} =
\begin{cases}
N_0+\frac{D{\phi\phi}}{4\Gamma} & \text{for } k=0,
\\
\frac{D_{\phi\phi}}{\Gamma}\left[ \frac{\mu^2+\Gamma^2}{E_{k}^2}+1 \right] & \text{for } k\neq 0,
\end{cases}
\end{equation}
which should be compared with Eq.~(\ref{tail}). The singularity at $k=0$ has been removed in Eq.~(\ref{nasz}), 
but an additional term $D_{\phi\phi}/4\Gamma$ indicates that on average the noise slightly increases the number of particles in the condensate.

For consistency, the size of the box must be sufficiently small so that the fraction of particles in the $k=0$ mode is considerable.
The number of particles in out-of-condensate modes can be estimated by replacing the sum of $n^{ss}_{\textbf{k}}$ over $k$ by an integral over the modes
from $\Delta k$ to $k_{\rm max}=2\pi/L$. From this estimate we get the self-consistency ``small box'' condition to be 
$L \ll \frac{4\pi^2\hbar |\phi_0|^2}{m D_{\phi\phi}} \frac{\Gamma \mu}{\Gamma^2+\mu^2}$ in the 1D case and
$L \ln \frac{\mu m L^2}{\hbar\pi^2}\ll \frac{8\pi^2\hbar |\phi_0|^2}{m D_{\phi\phi}} \frac{\Gamma \mu}{\Gamma^2+\mu^2}$ in the 2D case.

\section{Numerical results} \label{sec:numerical}

\begin{figure*}[tbp]
	\begin{center}
		\includegraphics[width=0.8\linewidth]{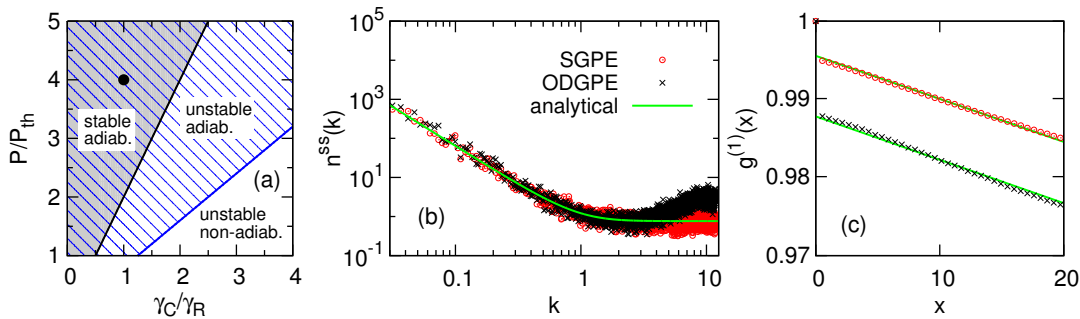} 
		\caption{(Color online).~ Stable and adiabatic regime. Left panel shows the position of the system on the 
stability-adiabaticity diagram. The black line corresponds to the stability condition~(\ref{condition}) and the blue line to the 
adiabaticity condition~(\ref{cond2}).
Middle panel shows momentum distribution in the steady state $n^{ss}(k)=|\psi(k)|^2 \Delta k$ for both models compared to the 
analytical prediction from~(\ref{tail}),~(\ref{nasz}). On the right panel the calculated first-order correlation functions $g^{(1)}(x)$ are displayed
together with the analytical long-range fit~(\ref{g1}).
Parameters in dimensionless ODGPE units~(\ref{dimensionless}) are $R = 1$, $g_C = 0.4$, $g_R = 2g_C$, $P/P_{\rm th}=4$, $\gamma_C = 1$, $\gamma_R = 1$,  $\beta=0.003$, $L=600$.}
		\label{fig:stab-adiab}
	\end{center}
\end{figure*}

\begin{figure*}[tbp]
	\begin{center}
		\includegraphics[width=0.8\linewidth]{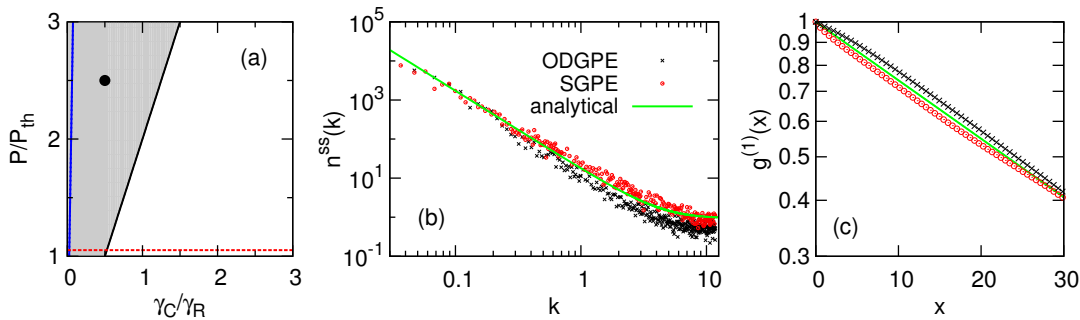} 
		\caption{(Color online).~ Stable and non-adiabatic regime. Same as in Fig.~\ref{fig:stab-adiab}, except for $g = 20$, $g_R = 2g$, $P/P_{\rm th}=2.5$, $\gamma_C = 0.5$.
}
		\label{fig:stab-nonadiab}
	\end{center}
\end{figure*}

In this subsection we investigate in detail the effects of the breakdown of the adiabatic approximation on the momentum distribution
and spatial correlation functions of a one-dimensional condensate. To this end we investigate numerically the steady states of the system
using both the 1D version of the SGPE model~(\ref{SGPE}) and the open-dissipative Gross-Pitaevskii model~(\ref{GPE-psi})-(\ref{GPE-nr}) with the
fluctuations included. We compare these numerical results with the analytical predictions given by~(\ref{tail}), (\ref{g1}) and (\ref{nasz}).

To investigate the breakdown of adiabaticity we consider various values of parameters for which the adiabaticity 
conditions~(\ref{cond1}) and~(\ref{cond2}) are either fulfilled or not. We clearly observe that the two models lead to the same results
only in the adiabatic regime. We also find that the discrepancy between the
SGPE and ODGPE models may be caused by large density fluctuations, which also invalidate the correspondence between the models.
This behaviour is observed close to the limits of condensate stability~(\ref{condition}) and $P=P_{\rm th}$.

Note that in the one-dimensional case the nonlinear coefficients scale with the confinement strength, 
eg.~$(R^{\rm 1D},g_i^{\rm 1D})=(R^{\rm 2D},g_i^{\rm 2D})/\sqrt{2\pi d^2}$, if we 
assume a Gaussian transverse profile of $|\psi|^2$ and $n_{\rm R}$ of width $d$.
In the case of a one-dimensional microwire~\cite{Bloch_ExtendedCondensates}, 
the profile width $d$ is of the order of the microwire thickness.
For convenience we also introduce dimensionless parameters in the system~(\ref{GPE-psi})-(\ref{GPE-nr}) 
which allows us to describe the system using a limited set of parameters.
By rescaling time, space, wave function amplitude and material coefficients as $t=\tau \widetilde{t}$, $x=\xi\widetilde{x}$, $\psi=(\xi\beta)^{-1/2}\widetilde{\psi}$, $n_R=(\xi\beta)^{-1}\widetilde{n}_R$  $R^{\rm 1D}=(\xi\beta/\tau)\widetilde{R}$, $(g^{\rm 1D},g_R^{\rm 1D})=(\hbar\xi\beta/\tau)(\widetilde{g},\widetilde{g}_R)$, $(\gamma_C,\gamma_R)=\tau^{-1}(\widetilde{\gamma_C},\widetilde{\gamma}_R)$, $P(x)=(1/\xi\beta\tau)\widetilde{P}(x)$, where $\xi=\sqrt{\hbar \tau /2m^*}$, while $\tau$ and $\beta$ are arbitrary scaling parameters, we can rewrite the above equation in the dimensionless form (from now on we omit the tildas for convenience)
\begin{align}\nonumber
&i\frac{\partial \psi}{\partial t}=\left[ -\frac{\partial^2}{{\partial x}^2}+\frac{i}{2}\left(Rn_R-\gamma_C\right)+g_C|\psi|^2+g_Rn_R\right]\psi,\\
\label{dimensionless}
&\frac{\partial n_R}{\partial t}=P-\left(\gamma_R+R|\psi|^2\right)n_R,
\end{align}
In particular, we may choose the time scaling $\tau$ in such a way that $\gamma_R=1$ without loss of generality. 
The norms of both fields $N_\psi=\int|\psi|^2 dx$ and $N_R=\int n_R dx$ are scaled by the factor of $\beta$,
and the noise correlators become $\langle dW(x) dW^*(x')\rangle = \beta\frac{dt}{2 \Delta x} (R n_R + \gamma_C) \delta_{x,x'}$.
The parameters of the SGPE model~(\ref{SGPE}) can be rescaled in an analogous way.

The numerical simulations were performed on a grid of length $L=600$ with spatial step size $\Delta x=0.133$. We solved the equation~(\ref{SGPE}) 
and~(\ref{GPE-psi})-(\ref{GPE-nr}) with the corresponding parameters given by the conditions~(\ref{correspondence}). We compared
the calculated momentum distribution $n^{ss}(k)$ and the first-order correlation function 
$g^{(1)}(x)$ in the steady state after long time evolution, starting 
from the initial state with $\phi(x,t=0)=\phi_0$ and $\nR(x,t=0)=\gamma_C/R$. We found that typically 
the result does not depend on the initial conditions. However, for certain values of
parameters, close to the stability threshold~(\ref{condition}), an empty initial state $\phi(x,0)=0$ 
led to a qualitatively different evolution (see the next subsection for details).

\begin{figure*}[tbp]
	\begin{center}
		\includegraphics[width=0.8\linewidth]{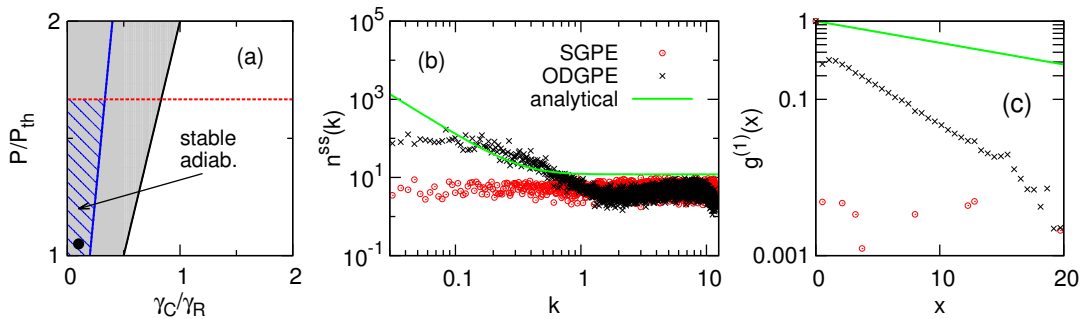} 
		\caption{(Color online).~Steady state close to the condensation threshold, exhibiting large fluctuations. Same as in Fig.~\ref{fig:stab-adiab}, except $g = 2.5$, $g_R = 2g$, $P/P_{\rm th}=1.05$, $\gamma_C = 0.1$.}
		\label{fig:Pth}
	\end{center}
\end{figure*}

\begin{figure}[tbp]
	\begin{center}
		\includegraphics[width=0.95\linewidth]{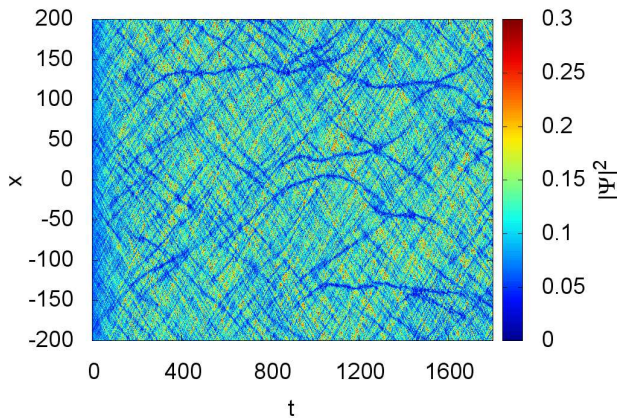} 
		\caption{(Color online).~Density evolution corresponding to Fig.~\ref{fig:Pth}, revealing dark solitons in the
dynamical steady state of the ODGPE.}
		\label{fig:solitons}
	\end{center}
\end{figure}

\subsection{Stable-adiabatic regime}

In Figure~\ref{fig:stab-adiab} we show the results in the stable-adiabatic case, i.e.~for parameters fulfilling both the stability~(\ref{condition})
and adiabaticity conditions~(\ref{cond1})-(\ref{cond2}). The left panel shows the position on the phase diagram. 
The middle panel displays the numerically obtained momentum distribution $n^{ss}(k)=|\psi(k)|^2 \Delta k$ from 15 simulations of both models,
compared with the analytical prediction of the momentum ``tail'' ($k>0$) of~(\ref{tail}) and~(\ref{nasz}).
While the numerical box size $L$ is so large that we are in a quasicondensate rather than 
condensate regime~\cite{Carusotto_NonequilibriumQuasicondensates}, and the fraction of particles at $k=0$
is very small, the momentum tail distribution closely follows the analytical predictions. This follows from the 
fact that in the quasicondensate regime the Bogoliubov approximation can still be applied to a slice of the system where the quasicondensate
phase is approximately constant. 

The discrepancy between the SGPE and ODGPE models is visible only at high momenta, where the first (momentum-dependent) adiabaticity
condition~(\ref{condk}) breaks down. Indeed, this condition gives the upper limit $k\ll 8$ , which coincides with the value of $k$ at which
the ODGPE results start to deviate from the SGPE and analytical predictions. We note that this limit corresponds to very high momentum values.

On the other hand, the $g^{(1)}(x)$ function follows the analytically predicted long-range trend~(\ref{g1}) very closely for both models. 
The difference
between the two is in the different value of the constant in front of the exponent in~(\ref{g1}), which can be again attributed to the 
breakdown of adiabaticity at very high momenta, related to short distances.

Further, we performed a series of numerical tests in the stable-adiabatic regime with other values of parameters. 
With increasing ratio $\gamma_C/\gamma_R$, moving towards the stability threshold~(\ref{condition}), the differences between 
analytical and numerical results become more notable. Beyond the stability threshold, which corresponds to $g_0=0$, the comparison is no longer
possible since the analytical formulas are valid only for the case of repulsive effective interactions, $g_0>0$.

\subsection{Stable-nonadiabatic regime}

Figure~\ref{fig:stab-nonadiab} shows the results in the stable-nonadiabatic case, when the adiabaticity conditions~(\ref{cond1}) and~(\ref{cond2})
are not fulfilled. In this case we increased the strength of the interactions $g_C$ and $g_R$.
There is a visible discrepancy between the models both in the momentum distribution and the first-order correlation function,
which marks the breakdown of the adiabatic approximation. Note that the results are presented in the logarithmic scale, and the
actual difference between the calculated averages of $n^{ss}(k)$ differ significantly. The SGPE result still follows the Bogoliubov
analytical prediction closely. The correlation function only slightly differs from the analytical prediction for both models, as shown 
in the right panel.

\begin{figure*}[tbp]
	\begin{center}
		\includegraphics[width=0.6\linewidth]{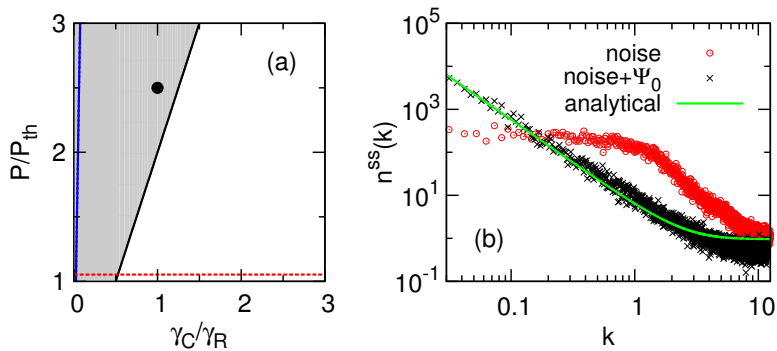} 
		\caption{(Color online).~System close to the stability threshold. 
In the right panel momentum distributions obtained from different initial conditions are shown (see text).
Parameters are as in Fig.~\ref{fig:stab-nonadiab}, except $\gamma_C = 1$.}
		\label{fig:treelets}
	\end{center}
\end{figure*}

\subsection{Large fluctuations} \label{sec:large_fluctuations}

Importantly, as shown in previous sections, the adiabaticity conditions~(\ref{condk})-(\ref{cond2}) 
are valid only under the assumption that the system is in a steady state with small fluctuations. Moreover, without this assumption
the reduction of the generalized CGLE~(\ref{CGLE-nr}) to the CGLE or SGPE models is not possible. We illustrate the situation 
in which the fluctuations are large in Fig.~\ref{fig:Pth}, where the pumping was chosen slightly above the threshold value $P\approx P_{\rm th}$.
In this case even the generalized Bogoliubov approximations are not relevant and the analytical predictions are incorrect. Moreover,
the results provided by the SGPE and ODGPE models are qualitatively different. The SGPE model predicts momentum distribution $n^{ss}(k)$
that is practically independent of momentum and negligible spatial correlations. On the other hand, the ODGPE predicts
a nontrivial momentum distribution and decaying $g^{(1)}(x)$ indicating an appearance of a finite spatial correlation length, not related to the
Bogoliubov prediction~(\ref{g1}). 

We associate this spatial length scale with the spontaneous appearance of dark structures depicted 
in Fig.~\ref{fig:solitons}. The figure shows density distribution in one randomly chosen realization of the truncated Wigner 
stochastic evolution~(\ref{GPE-psi})-(\ref{GPE-nr}), which can be interpreted as a single realization of the experiment~\cite{Wouters_ClassicalFields,Matuszewski_UniversalityPolaritons}. The structures appear to be related to dark solitons of the 
dissipative model~\cite{Xue_DarkSolitons,Flayac_DarkSolitons,Ostrovskaya_DarkSolitons}.
We checked that each dark object corresponds to an approximate $\pi$ phase jump of the phase of $\psi(x,t)$. 
A detailed investigation of these excitations will be presented elsewhere.


\subsection{Stability threshold and the role of initial conditions}

In the majority part of the stability diagram, the initial conditions given as a seed to the evolution do not influence the steady state
properties. However, the situation is drastically different for the ODGPE model in the vicinity of the stability threshold given by~(\ref{condition}). 
An example is given in Fig.~\ref{fig:treelets}. Here, the momentum distribution is plotted for simulations starting from two 
different initial conditions, either $\psi(x,t=0)=\psi_0+\xi(x)$ or $\psi(x,t=0)=\xi(x)$, where $\xi$ represents a small uncorrelated noise
with a Gaussian distribution. In the first case, the system converges to a steady state that is very well described by the analytical 
Bogoliubov momentum distribution. In the second case, the system does not reach this state even after very long evolution,
instead dwelling in a chaotic evolution with large density fluctuations. The ``normal'' and ``chaotic'' states are therefore metastable.
This behavior can be generally observed in the vicinity of the
stability limit, both on the stable and unstable side of the phase diagram of Fig.~\ref{fig:treelets}. 
We checked that this bistability persits even if a relaxation term (frequency-dependent pumping) 
is included the evolution equation (\ref{GPE-psi}), analogous as in~\cite{Wouters_CriticalVelocity}. Since we do not find
any similar dynamics in the SGPE model, we conclude that it is also related to the breakdown of the adiabatic approximation. 
The investigation of these chaotic states will be a topic of a future study.

\section{Conclusions} \label{sec:conclusions}

In conclusion, we investigated the relation between the  models commonly used in the literature 
to describe nonresonantly pumped exciton-polariton condensates. The complex Ginzburg-Landau equation, 
and the equivalent (in the limit of small fluctuations) stochastic Gross-Pitaevskii equation were
compared with the open-dissipative Gross-Pitaevskii equation 
which includes a separate equation for the reservoir density.
The adiabatic approximation allows to reduce the latter to one of the single-equation models, 
under the assumptions that the condensate is close to the steady state and fluctuations are small.
Additionally, three independent analytical conditions for adiabaticity must be met simultaneously. 
While spin degree of 
freedom was not taken into account in this work, the generalization to the spin-dependent case is 
straightforward.

We investigated the corresponding stochastic models by comparing the numerical
steady states with the analytical predictions of the Bogoliubov approximation. 
We demonstrated how the zero-momentum singularity of the momentum distribution spectrum can be avoided by
an appropriate generalization of the Bogoliubov approximation. This also allowed for determination of
the number fluctuations and condensate phase diffusion equation.

The comparison of the models with and without a separate equation for the reservoir 
demonstrated that close agreement between the two can be obtained only under the adiabatic conditions.
Moreover, we showed that close to the limit of condensation or the limit of modulational stability,
large fluctuations lead to qualitatively different results depending on the model used.
These results show that special care must be taken when choosing the right model for describing 
exciton-polariton condensates in certain conditions.

\acknowledgments

We thank Iacopo Carusotto for reading the manuscript and useful comments.
This work was supported by the National Science Center grant DEC-2011/01/D/ST3/00482. 

\bibliography{references}
\end{document}